\documentstyle [12pt] {article}
\title{IS OPERA NEUTRINO SUPERLUMINAL PROPAGATION SIMILAR TO GAIN-ASSISTED SUPERLUMINAL LIGHT PROPAGATION}

\author{Vladan Pankovi\'c,\\
Department of Physics, Faculty of Sciences, 21000 Novi Sad,\\ Trg
Dositeja Obradovi\'ca 4, Serbia, \\vladan.pankovic@df.uns.ac.rs}

\date {}
\begin {document}
\maketitle \vspace {0.5cm}

\begin {abstract}
In this work we consider a possible conceptual similarity between
recent, amazing OPERA experiment of the superluminal propagation
of neutrino and experiment of the gain-assisted superluminal light
propagation realized about ten years ago. Last experiment refers
on the propagation of the light, precisely laser pulse through a
medium, precisely caesium atomic gas, with characteristic
anomalous dispersion and corresponding negative group-velocity
index with very large amplitude between two closely spaced gain
lines (that is in some way similar to quantum theory of the
ferromagnetism). It implies superluminal propagation of the light
through this medium. Nevertheless all this, at it has been pointed
out by authors, "is not at odds with causality or special
relativity", since it simply represents "a direct consequence of
the classical interference between … different frequency
components". We suggest that OPERA experiment can be in some way
conceptually similar to the gain-assisted superluminal light
propagation experiment. For this reason we suppose too that OPERA
experiment can be simply explained in full agreement with
causality and special relativity if there is some medium,
precisely a scalar field (e.g. dark matter field, Higgs field or
similar) through which neutrino propagates. We prove that,
according to OPERA experiment data, supposed medium must be
non-dispersive while its refractive index must be positive,
smaller but relatively close to 1 (that is in some way similar to
quantum theory of the diamagnetism). If it is true OPERA
experiment results do not mean that special theory of relativity
is broken, but they mean detection of suggested medium, i.e. a
scalar field (e.g. dark matter field, Higgs field or similar).
\end {abstract}

\vspace {0.5cm} PACS number: 12.15.-y, 13.15.-g \vspace {0.5cm}

In this work we shall consider a possible conceptual similarity
between recent, amazing OPERA experiment of the superluminal
propagation of neutrino [1] and experiment of the gain-assisted
superluminal light propagation [2] realized about ten years ago.
Last experiment refers on the propagation of the light, precisely
laser pulse through a medium, precisely caesium atomic gas, with
characteristic anomalous dispersion and corresponding negative
group-velocity index (very large amplitude between two closely
spaced gain lines) that implies superluminal propagation of the
light through this medium. (It is in some way similar to quantum
theory of the ferromagnetism according to which relative magnetic
permeability becomes large in a narrow boundary between two
domains.) Nevertheless all this, at it has been pointed out by
authors, "is not at odds with causality or special relativity",
since it simply represents "a direct consequence of the classical
interference between … different frequency components". We shall
suggest that OPERA experiment can be conceptually similar to the
gain-assisted superluminal light propagation. For this reason we
shall suppose that OPERA experiment can be simply explained in
full agreement with causality and special relativity if there is
some medium (a scalar field, e.g. dark matter field, Higgs field
or similar) through which neutrino propagates. We shall prove
that, according to OPERA experiment data, supposed medium must be
non-dispersive while its refractive index must be positive,
smaller but relatively close to 1. (It is in some way similar to
quantum theory of the diamagnetism which needs that orbital
momentum of the diamagnetic atoms must be exactly equivalent to
zero and which predicts that relative magnetic permeability is
smaller but relatively close to 1.) If it is true OPERA experiment
results do not mean that special theory of relativity is broken,
but they mean detection of suggested medium, i.e. a scalar field
(e.g. dark matter field, Higgs field or similar).

About ten years ago Wang, Kuzmich and Dogariu [2] realized
experiment of the gain-assisted superluminal light propagation
whose basic elements (including corresponding theory) we shall now
shortly repeat. As it is well-known in a dispersive linear medium
with optical refractive index $n(\nu)$ depending of the optical
frequency $\nu$ light pulse with this frequency propagates with
the group velocity $v_{g}= \frac {c}{n_{g}}$  where $n_{g}= n(\nu)
+ \nu \frac {dn(\nu)}{d\nu}$ represents the group-velocity
refractive index and $c=299792$km/s - speed of light. In domain
between two closely spaced gain lines there is an anomalous
dispersion region where $\nu \frac {dn(\nu)}{d\nu}$ is negative
with extremely large amplitude. (It is in some way similar to
quantum theory of the ferromagnetism according to which relative
magnetic permeability becomes large in a close space between two
domains.) In this situation expression
\begin {equation}
    \Delta T = \frac{L}{v_{g}}- \frac {L}{c} = (n_{g} -1) \frac {L}{c}
\end {equation}
that represents time difference between propagation time of the
light pulse through medium with length $L$ and propagation time of
the light pulse through vacuum with the same length $L$, becomes
negative too. It means that light pulse propagates through medium
effectively superluminally, i.e. faster than propagation of this
pulse through vacuum (pulse time advancement shift). Nevertheless
all this, at it has been pointed out by authors, "is not at odds
with causality or special relativity", since it simply represents
"a direct consequence of the classical interference between …
different frequency components" [2]. "Remarkably, the signal
velocity of a light pulse, defined as the velocity at which half
point of the pulse front travels, also exceed the speed of light
in vacuum, $c$, in present experiment. It has also been suggested
that the true speed at which information is carried by a light
pulse should be defined as the "frontal" velocity of the
step-function-shaped signal which has been shown not to exceed
$c$."[2] In experimental realization of this theory Wang, Kuzmich
and Dogariu used gaseous medium of the caesium atoms any of which
has one excited state and two (close) ground states and
corresponding polarized laser beam and obtained pulse advancement
shift $\Delta T = 62 ns$ or $n_{g} = -310 $ for $L = 6 cm$.

In recent OPERA experiment [1], in agreement with some other
earlier experiments on the superluminal neutrino propagations [3],
there is a pulse of muon neutrinos that propagates along base line
\begin {equation}
     L=730534.61 m
\end {equation}
, with time difference with respect to the one assuming the speed
of light in vacuum
\begin {equation}
     \Delta T = -60.7 ns
\end {equation}
or with relative difference of the muon neutrino velocity v with
respect to the speed of light
\begin {equation}
    \frac {v-c}{c}=2.48 10^{-5}
\end {equation}
corresponding to
\begin {equation}
     v = (1 + 2.48 10^{-5}) c = 1.0000248 c .
\end {equation}

It represents an extremely unexpected result whose theoretical
explanation in this moment is unknown. For example Amelino-Camelia
group [4] supposes that OPERA data can be explained by
special-relativistic tachyons, etc.

We shall originally and simply suppose that OPERA experiment is in
some way conceptually similar to the experiment of the
gain-assisted superluminal light propagation. Really, many
characteristics of the neutrinos are very similar to the
characteristic of the photons. But it can be observed that in the
experiment of the gain-assisted superluminal light propagation
there is characteristic medium, precisely caesium atomic gas, with
characteristic anomalous dispersion and corresponding negative
group-velocity index, while in the OPERA experiment similar
medium, at the first sight, does not exist. However, it can be
observed that a scalar field, e.g. dark matter field, Higgs field
or similar, can exist. This field seems practically identical to
any observer, moving or rest, and mimics vacuum. Moreover, as it
has been pointed out by Linde (in the chaotic inflation cosmology
[5], [6]) such field can during time occupy practically whole
space or, at least, our galaxy, Sun system and Earth. Finally, a
quantized scalar field has quantums with zero spin representing
bosons.

If such suggested scalar field really exists and if it has some
group velocity refractive index for neutrino, this refractive
index in OPERA experiment, according to (1), (2), (3), equals
\begin {equation}
   n_{g} = 0.975                                        .
\end {equation}
It represents a positive refractive index smaller but relatively
close to 1. All this is in some way similar to quantum theory of
the diamagnetism which predicts that relative magnetic
permeability is smaller but relatively close to 1 and  which needs
that orbital momentum of the diamagnetic atoms must be exactly
equivalent to zero. If this similarity has any sense it would mean
that quantums of mentioned scalar field hold zero spin in full
agreement with general quantum theory of the scalar fields.

Also, according to OPERA experiment data, there is no dependence
between $\Delta T$ or $ n_{g}$ and neutrino energy or (de Broglie)
frequency, which means that suggested medium must be
non-dispersive.

Finally, it can be observed that introduced hypothesis on the
scalar field representing non-dispersive medium for neutrinos,
admits that, like in case of the gain-assisted superluminal light
propagation, superluminal propagation of the neutrinos be
explained in simple way that "is not at odds with causality or
special relativity", since it simply represents, we paraphrase, "a
direct consequence of the wave characteristics" of neutrino.

In conclusion we shall shortly repeat and point out the following.
In this work we consider a possible conceptual similarity between
recent, amazing OPERA experiment of the superluminal propagation
of neutrino and experiment of the gain-assisted superluminal light
propagation realized about ten years ago. Last experiment refers
on the propagation of the light, precisely laser pulse through a
medium, precisely caesium atomic gas, with characteristic
anomalous dispersion and corresponding negative group-velocity
index (with very large amplitude between two closely spaced gain
lines) that implies superluminal propagation of the light through
this medium. (It is in some way similar to quantum theory of the
ferromagnetism according to which relative magnetic permeability
becomes large in a narrow boundary between two domains.)
Nevertheless all this, at it has been pointed out by authors, "is
not at odds with causality or special relativity", since it simply
represents "a direct consequence of the classical interference
between … different frequency components". We suggest that OPERA
experiment can be in some way conceptually similar to the
gain-assisted superluminal light propagation. For this reason we
suppose too that OPERA experiment can be simply explained in full
agreement with causality and special relativity if there is some
medium, precisely a scalar field (e.g. dark matter field, Higgs
field or similar) through which neutrino propagates. We prove
that, according to OPERA experiment data, supposed medium must be
non-dispersive while its refractive index must be positive,
smaller but relatively close to 1. (It is in some way similar to
quantum theory of the diamagnetism which needs that orbital
momentum of the diamagnetic atoms must be exactly equivalent to
zero and which predicts that relative magnetic permeability is
smaller but relatively close to 1.) If it is true OPERA experiment
results do not mean that special theory of relativity is broken,
but they mean detection of suggested medium, i.e. a scalar field
(e.g. dark matter field, Higgs field or similar).

\vspace{1cm}

{\large \bf References}

\begin{itemize}

\item [[1]]  T. Adam, N. Agafonova, A. Aleksandrov et al, {\it Measurement of the Neutrino Velocity with the OPERA Detector in the CNGS Beam}, hep-ex/1109.4897
\item [[2]] L. J. Wang, A. Kuzmich, A. Dogariu, Nature {\bf 406} (2000), 277
\item [[3]] P. Adamson et al, [MINOS collaboration], Phys. Rev. {\bf D76} (2007) 072005; hep-ex/0706.0437
\item [[4]] G. Amelino-Camelia et al, {\it OPERA-reassessing Data on the Energy Dependence of Speed of Neutrinos}, hep-ph/1109.5172
\item [[5]] A. D. Linde, in, {\it 300 Years of Gravitation}, eds. S. W. Hawking and W. Israel (Cambridge University Press, Cambridge, England, 1989)
\item [[6]] A. D. Linde, {\it Inflation and Quantum Cosmology} (Academic Press, Boston, 1990)

\end {itemize}

\end {document}